\documentclass[twocolumn,prc]{revtex4}
\usepackage{epsfig,graphicx,float,pst-all}
\usepackage{booktabs}
\usepackage{amssymb, amsmath}

\newcommand{\be}{\begin{equation}} 
\newcommand{\ee}{\end{equation}}
\newcommand{\bc}{\begin{center}}
\newcommand{\ec}{\end{center}}

\newcommand{\WD}{{\cal{D}}}

\usepackage{booktabs}

\begin{document}

\title{Transition densities and form factors in the triangular $\alpha$-cluster model of $^{12}$C with application to $^{12}$C+$\alpha$ scattering.}
\author{A.Vitturi$^{1,2}$, J. Casal$^{1,2}$, L.Fortunato$^{1,2}$ and E.G. Lanza$^{3,4}$ }
\affiliation{1) Dipartimento di Fisica e Astronomia ``G.Galilei'' - Universit\`a di Padova \\
2) I.N.F.N. - Sez. di Padova, via F.Marzolo 8,  I-35131 Padova, Italy \\
3) I.N.F.N. - Sez. di Catania, via S. Sofia 64, I-95123 Catania, Italy  \\ 
4) Dipartimento di Fisica e di Astronomia ``Ettore Majorana", Univ. Catania, Italy}

\begin{abstract}
	Densities and transition densities are computed in an equilateral triangular alpha-cluster model
	for $^{12}$C, in which each $\alpha$ particle is taken as a gaussian density distribution.
	The ground-state, the symmetric vibration (Hoyle state) and the asymmetric bend vibration are analyzed in a molecular approach and dissected into their components in a series of harmonic functions, revealing their intrinsic structures. The transition densities in the laboratory frame are then used to construct form-factors and to compute DWBA inelastic cross-sections for the $^{12}$C$(\alpha, \alpha')$ reaction. The comparison with experimental data indicates that the simple geometrical model with rotations and vibrations gives a reliable description of reactions where $\alpha$-cluster degrees of freedom are involved.
 
\end{abstract}
%\pacs{}
\maketitle

\section{Introduction}

Few nuclear systems have attracted the interest of the scientific community as the $^{12}$C nucleus that is remarkable under many respects: it is the nucleus at the center of the atom mostly associated with life on earth and yet it is produced in the billion-kelvins hot plasma of stars.
It is a crucial $N=Z$ even-even system that shows an unusual energy spectrum and rotational bands and despite the large number of experiments, various theoretical interpretations of data and a handful of models of its nuclear structure, it escapes conventional descriptions such as the single-particle shell model or the collective model. In particular, it still largely baffles the efforts of pinning its structure down with {\it ab initio} nuclear shell model based on realistic interactions, due to the strong tendency of nucleons to form clusters of alpha particles. This is a signal that even the most up-to-date NN interactions still miss some important ingredient that could explain clusterization in light nuclei. Reactions involving $^{12}$C as a target or as a projectile have been performed extensively due to the easiness to chemically or physically deal with this abundant isotope in order to produce targets or to build ion sources that can deliver intense ion beams. The acceleration of these ions to various energies allows the analysis of several types of reactions that highlight different aspects worth of interest, like cluster transfer reactions, nuclear rainbow
and a number of studies aimed at measuring the nuclear S-factor that is relevant for astrophysics. 

A large number of models have been constructed along the years with several degrees of success that cover various aspects of its complex phenomenology, but a final word has not yet been written \cite{Freer, Betts, Fre14}.
Very interestingly, instead of going into the direction of treating the system of $A=12$ particles with individual degrees of freedom, another line of investigation has been recently pursued, the Algebraic Cluster Model \cite{Bij, Bij02,Bij14, Stel, Stel2, Del17,Del17b, Bij95, MLB14}, in which a simplification in terms of rotational and vibrational excitations of an equilateral triangular configuration of three alpha particles seems to offer a valid explanation of most of the low-energy spectral features of $^{12}$C. This kind of molecular models have a long history, starting with Wheeler in 1937 \cite{Whe37,Whe37bis, Wefel, HTe38} and have been forgot or misunderstood in the past and left behind, in favor of fully microscopic approaches. Notably the criticisms contained in the book by Blatt and Weisskopf \cite{BlaWei}, which were certainly well-armed, but have been surpassed by new accumulated knowledge by now, have contributed to belittle this approach. On the other hand, it is true, anyway, that light nuclei, those who bridge the gap from deuterium to the mass region where the single-particle shell model starts to work beyond any doubts, have been subject to profound investigations both from the theoretical and experimental sides that support a molecular-like interpretation. A very important experimental work with a strong theoretical foundation was done by W. von Oertzen and coworkers, that have identified several molecular structures and rotational bands with certainty \cite{vonOe, vonOe2, vonOe3, Milin} and have formulated a molecular orbital theory, inspired to quantum chemical models, for the bonds of additional neutrons that move in the force fields of alpha-cluster structures. These structures, quite similarly to what happens in molecules, can vibrate and rotate around fixed positions, but quite differently to what happens in molecules, are not rigid at all. They rather are soft dynamical nuclear systems, whose zero point motion has the same size of the nucleus itself, therefore the underlying geometric configurations should be attributed to equilibrium points around which large fluctuations occur (in other words the Born-Oppenheimer approximation does not work here).
Rotations and vibrations have approximately the same energy scales and they are intertwined with stronger ties.

Linear chains of alpha particles \cite{Mor} as well as BEC gas of alpha bosons \cite{Toh} have been proposed as possible explanations of certain states or parts of the energy spectrum. The literature is rich with theoretical interpretation: cluster models (e.g.,~\cite{Nguyen13}), No-Core Shell Model (e.g.,~\cite{Drey, Nav}), as well as Antisymmetrized Molecular Dynamics (e.g.,~\cite{AMD}), Fermion Molecular Dynamics \cite{FMD,Chernykh07}, Effective Field Theory \cite{EFT} and lattice calculations (e.g.,~\cite{Lattice}), Non-localized clustering \cite{Zhou}, each with its own merits. A lucid analysis of all these models \cite{Freer} reveals however that no final agreement can be reached for the underlying geometric structure. Recently, one of us \cite{Fort} proposed a theoretical scheme based on discrete point-group symmetries, by which a Raman fluorescence experiment in which the depolarization ratio is measured for the excited states, might discriminate among all possible types of geometric configurations with certainty, thus offering a solution to this conundrum.

We have dealt with nuclear structure up to now, but reactions of $^{12}$C are of paramount importance, because, leaving aside the interest in understanding and modeling the reaction mechanisms themselves, it is through dynamical processes like collisions and absorption or emission of electromagnetic waves that we have a handle on the structural features, from which one could ultimately derive fundamental information on the nuclear interactions. Thus, it appears to us that some blending of these two wide chapters of physics has to be done and our chosen method will be that of transition densities. 
One of the earliest works in this respect, and very similar in spirit to ours, though based on a resonating group method calculation, was that of Kamimura \cite{Kam}. 
A folding-model analysis of the inelastic $^{12}$C$+\alpha$ scattering has been performed in Ref. \cite{Kan} using Antisymmetrized Molecular Dynamics to calculate wavefunctions and transition densities and either the Distorted Wave Born Approximation (DWBA) or Coupled channels to compute the differential cross-sections. 
More recently, Ito \cite{Ito} linked the coupled-channel fit of inelastic scattering data to the extended nature of the $2_2^+$ state in the Hoyle band.
Even more recently, Kanada-En'yo and Ogata
came up with a reanalysis of the $\alpha$ scattering cross sections in a coupled-channel formalism  \cite{KanOg}. In this reference the monopole and dipole excitations and several other observables are discussed and in the conclusions it is stated that further calculations are required to reduce the ambiguities of several parameters entering the structure calculations. 

We have given a preliminary account of some of the feature of our approach in Ref. \cite{Vitturi19}. Therefore in the present paper we will proceed to investigate various aspects of structure and reactions in connection with the occurrence of alpha clusters in the $^{12}$C nucleus. We will begin by studying the equilateral triangle model, its density and transition densities not only for the ground state band, but also for excited vibrational bands.  
Then we will use it to calculate form factors between these states and these, in turn, will be used to compute inelastic scattering cross-sections in DWBA.
Our main aim is to show that a simple description in terms of rotations and vibrations of triangular configurations is sufficient to yield all the relevant features of the inelastic process. While complicated models including nucleon-nucleon interactions are certainly more advanced, they are not necessary to describe the most salient features of this process. A simple model based on symmetry accounts for practically all of the relevant facts.

\section{Densities and transition densities}
The density of the $\alpha$ particle is taken as a gaussian function:
\be
\rho_\alpha(\vec r)= \Bigr(\frac{\alpha}{\pi}\Bigl)^{3/2} e^{-\alpha r^2}
\ee
with $\alpha=0.56(2)$ fm$^{-2}$ as in Ref. \cite{Del17, Del17b}. 
The three dimensional spherical integral of this function is normalized to 1, therefore one should always multiply by 2 (the charge of an alpha particle), when dealing with charge-related quantities and multiply by 4 (the mass of an alpha), when dealing with mass-related properties.
Now, with the aim of constructing the density of $^{12}$C as a sum of three $\alpha$ particles placed at the vertices of a triangle, each particle should be displaced of the proper amount, $\beta$, therefore we have  
\be
\rho_{gs}(\vec r, \{ \vec r_k\})= \sum_{k=1}^3 \rho_\alpha(\vec r - \vec r_k)
\label{dez}
\ee
with $\vec r_1=(\beta,\pi/2,0)$, $\vec r_2=(\beta,\pi/2,2\pi/3)$ and  $\vec r_3=(\beta,\pi/2,4\pi/3)$ in spherical polar coordinates $(r,\theta,\phi)$, where the co-latitude is always $\pi/2$ because we have chosen a triangle lying in the $\{xy\}$ plane with the particle labeled as $1$ on the positive $x-$axis. Once the angular position of the alpha particles has been decided, the dependence on the 9 variables contained in the three vectos $ \{ \vec r_k\}$ is reduced to the single radial variable that we have called $\beta$.
With the proviso made above, the integral of this density is normalized to 3, therefore, once again, one should properly multiply by 2 or 4 depending on what is the aim of the calculations.
The shape of this `static' ground-state density, labeled with $gs$, is associated with the fully symmetric representation, $A$, of $D_{3h}$ with $0$ quanta of excitation \cite{Stel, Stel2}.
In the following, $\beta$ will be set to a constant value, therefore the explicit dependence on it can be dropped. Thus the density can be expanded in spherical harmonics as
\be
\rho_{gs}(\vec r)= \sum_{\lambda\mu} \rho_{gs}^{\lambda,\mu} (r) Y_{\lambda,\mu} (\theta,\varphi)
\ee
where  $\rho_{gs}^{\lambda,\mu}$ are the intrinsic radial transition densities that depend on $\lambda,\mu$. Our choice of coordinates is such that only those multipoles that are allowed by the $D_{3h}$ symmetry appear in the sum. This is different from Ref. \cite{Del17} where the $z-$axis was instead chosen to pass through particle $1$ and the center of the triangle. Once the densities are known in the intrinsic frame, they should be transformed into the laboratory frame, where the dependence on $\mu$ is lost. Details on how to accomplish this are given in the appendix.

The lab-frame radial transition densities allow the calculation of several intra-band observables, such as the reduced electromagnetic transitions $B(E\lambda)$ in terms of the corresponding matrix elements $M(E\lambda)$ defined as:
\begin{equation}
M(E2; 2^+_1 \rightarrow 0^+_1)= Z \int  \rho_{gs}^{\lambda =2} (r) r^4 dr
\end{equation}
where $Z=2$ is the charge of a single $\alpha$ particle and more in general
\begin{equation}
M(E\lambda; \lambda \rightarrow 0^+_1)= Z \int  \rho_{gs}^{\lambda} (r) r^{(\lambda+2)} dr
\end{equation}
\begin{equation}
B(E\lambda; \lambda \rightarrow 0^+_1)= \frac{1}{2\lambda +1}\mid M(E\lambda; \lambda \rightarrow 0^+_1) \mid ^2
\end{equation}
and the diagonal matrix elements and root mean square radius defined as:
\begin{equation}
M(E0) =\sqrt{4\pi} Z \int \rho_{gs}^{0} (r) r^4 dr
\end{equation}
\begin{equation}
\langle r^2 \rangle^{1/2}_{0^+_1} =\Bigl( \sqrt{4 \pi} \int  \rho_{gs}^{0} (r) r^4 dr /{\cal N} \Bigr)^{1/2}
\end{equation}
where ${\cal N} =  \sqrt{4 \pi}\int  \rho_{gs}^{0} (r) r^2 dr =3 $ is a normalization integral that just counts the number of $\alpha$ particles.

\begin{figure}[!t]
\includegraphics[width=.54\textwidth, clip=]{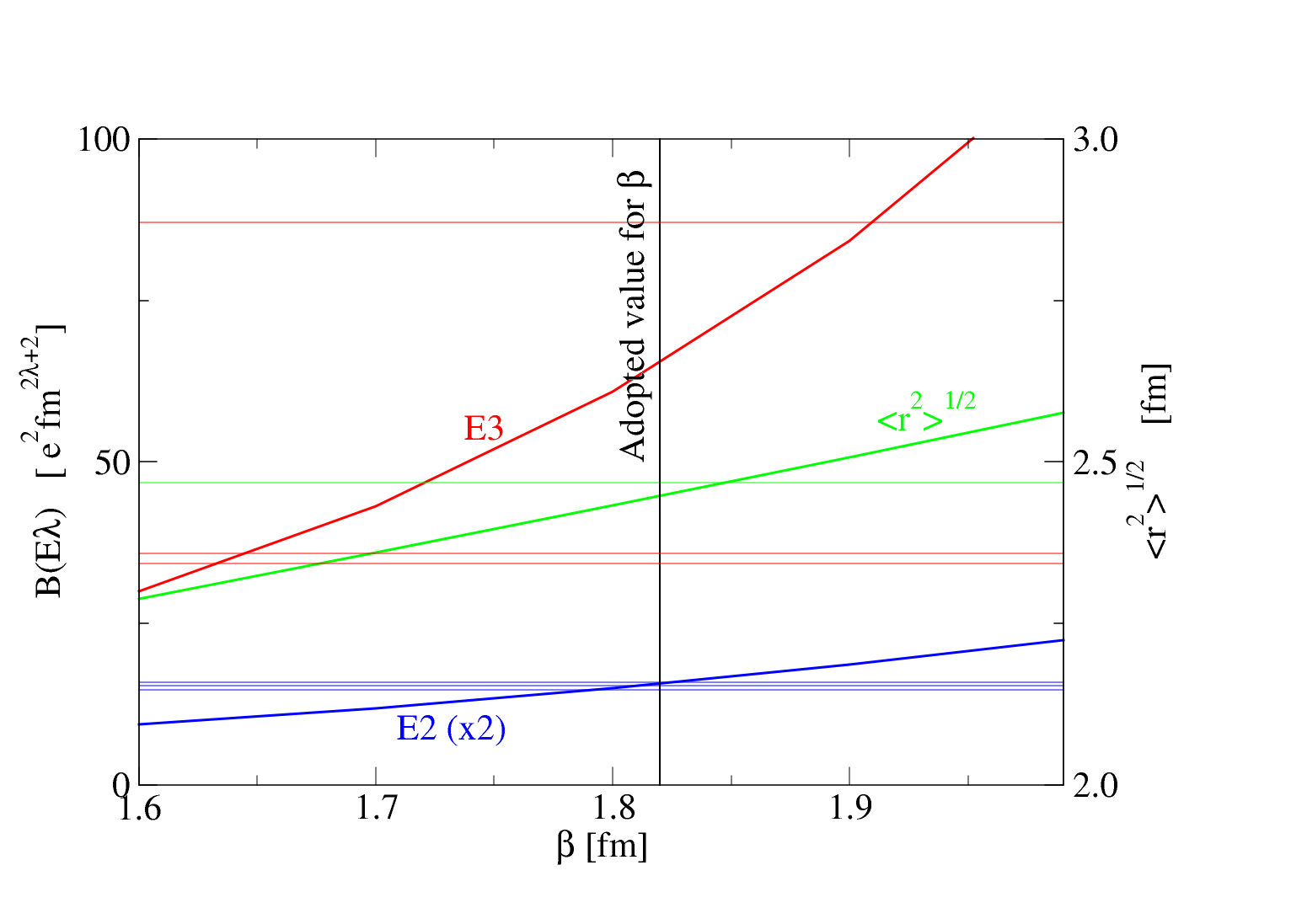}
\caption{Calculated values of $B(E2; 2^+_1 \rightarrow 0^+_1)$ and $B(E3; 3^+_1 \rightarrow 0^+_1)$ are shown as blue and red thick lines, while experimental values (without error bands) are shown as light horizontal lines of the same color. Units are on the left vertical axis, notice that the E2 has been multiplied by two. We plot in green the r.m.s. radius and only one experimental value for reference (units on the right vertical axis). The black vertical line at $\beta=1.82$ marks the value adopted in this paper.} 
\label{data0}
\end{figure}

Now one might take the radial parameter for distance as $\beta=1.74(4)$ fm for $k=3$ clusters as in Ref. \cite{Del17,Del17b}, but we prefer to choose $\beta=1.82$ fm because this allows to fix both the ground state radius and the $B(E2)$ to the first excited $2^+$ state. The change of calculated values with $\beta$ is compared with available experimental data in Fig. \ref{data0}. Horizontal coloured lines are measurements, while the vertical black line is the adopted value, i.e. a compromise between $B(E2)$ and the root mean square radius. Experimental values for $B(E3)$ are either too small or too large and in any case they do not agree with each other, but our adopted value falls in the middle. 

\begin{figure}[!t]
\includegraphics[width=.49\textwidth, clip=]{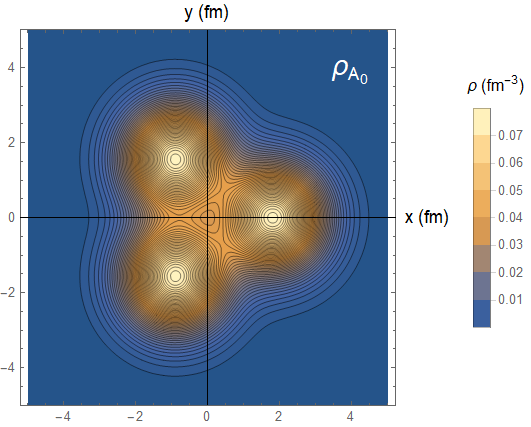}
\caption{Contour plot of density in fm$^{-3}$ (cut on the $z=0$ plane), $\rho_{gs}$ in Eq. \ref{dez}, of the g.s. static triangular configuration (with A symmetry).}
\label{cp0}
\end{figure}

\begin{figure}[!t]
	\includegraphics[width=1.\columnwidth, clip=]{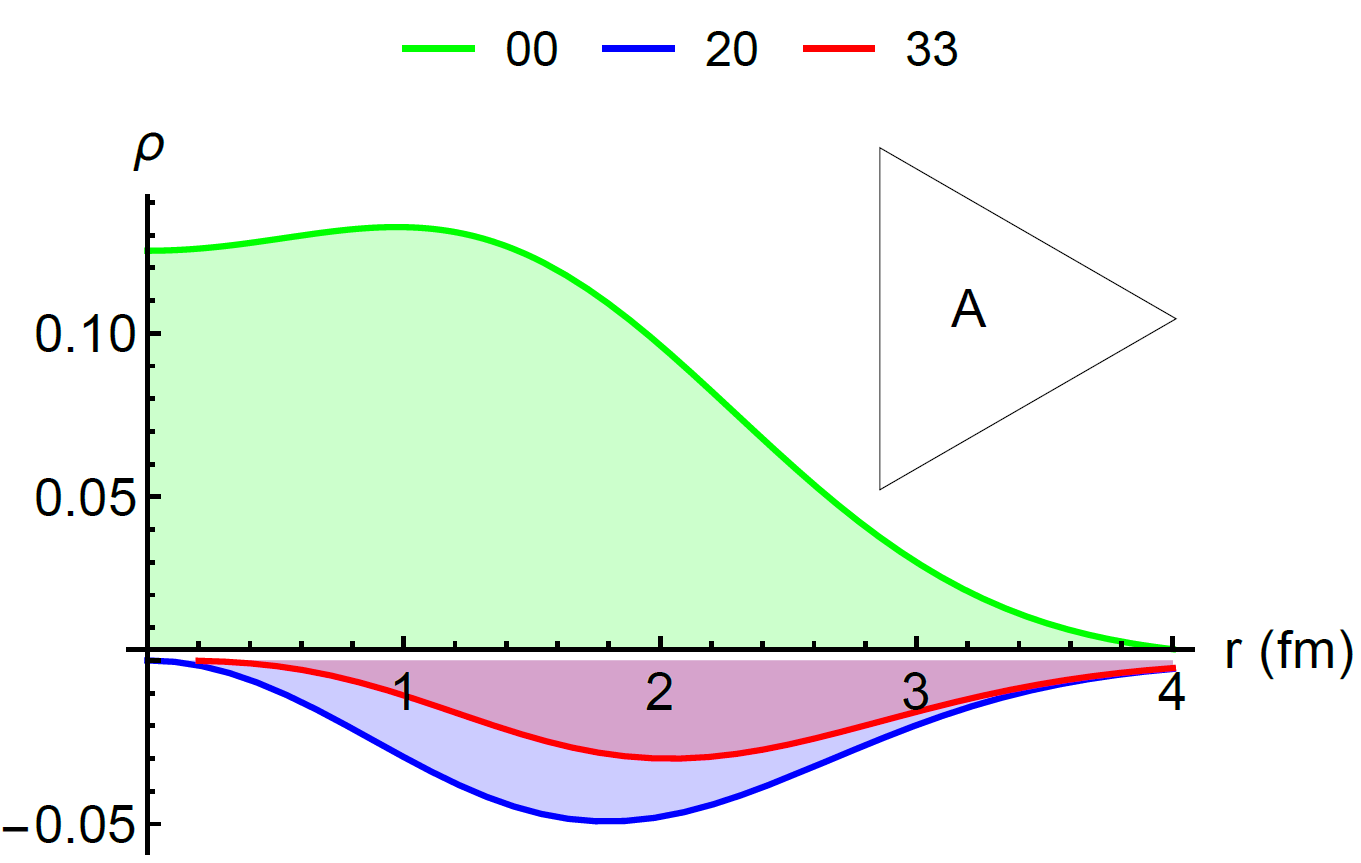}
	\caption{Radial transition densities, $\rho_{gs}^{\lambda,\mu}$ in Eq. \ref{dez}, of g.s. band with A symmetry.}
	\label{td0}
\end{figure}

Returning now to the density defined above,
Fig. \ref{cp0} shows a contour plot of the static triangular configuration associated with the ground-state band, while Fig. \ref{td0} shows the three lowest order radial functions of the expansion in spherical harmonics for $\{\lambda\mu\}=\{00,20,33\}$. The function labeled $00$ represents the ground state density while the others represent the change in density for transitions to higher lying states of the ground state band.
The properties of the g.s. band can be derived from the knowledge of the transition densities.
We collect in Table \ref{Ta1} the calculated values of r.m.s. radius, $B(E2)$, and $B(E3)$ obtained with $\beta =1.82$ fm.
\begin{table}[!b]
	\caption{Calculated observables within the g.s. band.}	\label{Ta1}
	\begin{tabular}{lll}\hline\hline
		$\langle r^2 \rangle^{1/2}_{0^+_1}$  &~& 2.45 (fm)\\ \colrule
		$B(E2; 2^+_1 \rightarrow 0^+_1) $ &~& 7.86 (e$^2$fm$^4$)\\ \colrule
		$B(E3; 3^-_1 \rightarrow 0^+_1) $ &~& 65.07 (e$^2$fm$^6$)\\\colrule
		$B(E4; 4^+_1 \rightarrow 0^+_1) $ &~& 96.99 (e$^2$fm$^8$)\\ \botrule
	\end{tabular}
\end{table}

\begin{figure}[!t]
	\includegraphics[width=1.\columnwidth, clip=]{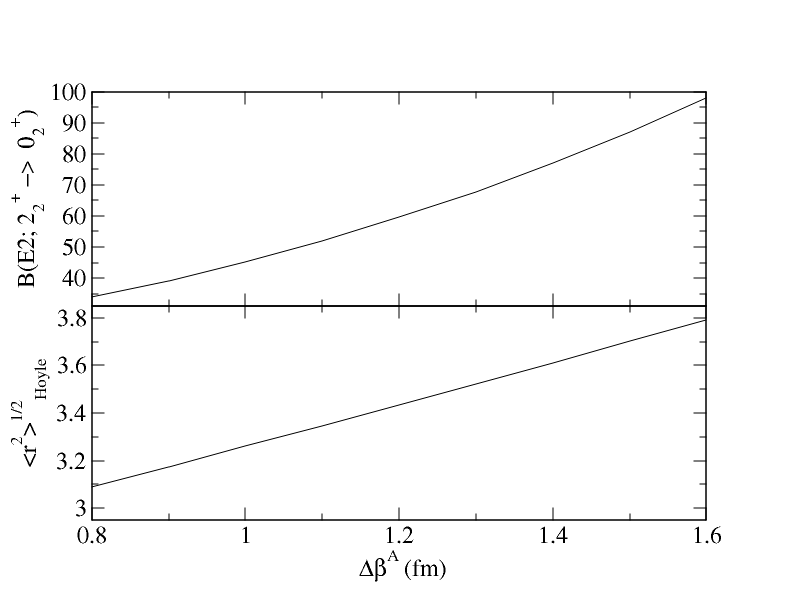}
	\caption{Calculated values of  $B(E2; 2^+_2 \rightarrow 0^+_2)$ (upper panel in $e^2$ fm$^4$) and $\langle r^2\rangle^2_{0^+_2}$ (lower panel, in fm) as a function of $\Delta \beta^A$.}
	\label{bandaA}
\end{figure}

\begin{figure}[!t]
	\includegraphics[width=1.\columnwidth, clip=]{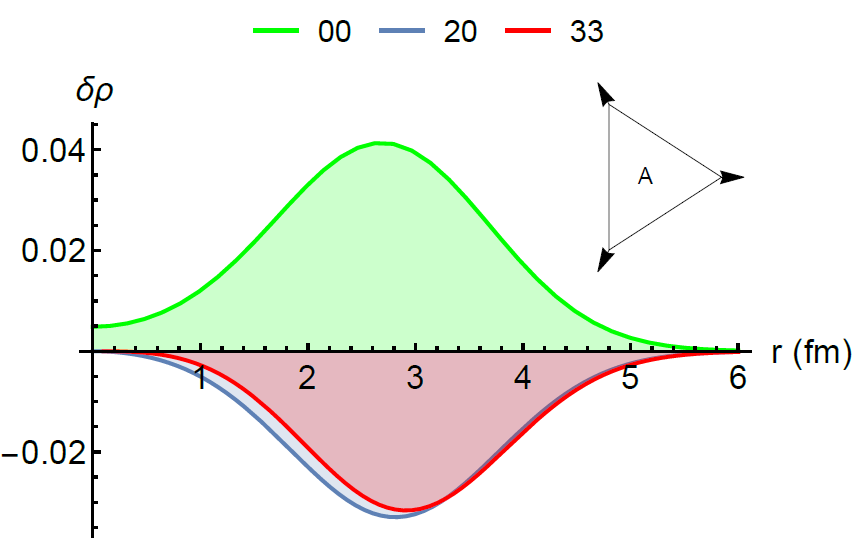}
	\caption{Radial transition densities, $\delta \rho_{gs \rightarrow A }^{\lambda,\mu}(r)$ of Eq. \ref{expdrho}, within the excited A-band.}
	\label{rd1}
\end{figure}

It is very important also to investigate what happens when the particles are displaced of a small amount along the directions of the vectors of normal modes of motion, that are of two types: singly-degenerate fully-symmetric, $A$, and doubly-degenerate, $E$. For example we can obtain the first symmetric vibrational band of $A-$type  (with $n=1$) by adding a small displacement  $\Delta \beta^A$ along the arrows of the inset in Fig. \ref{rd1}. It amounts to
redefine the radial variable in $\vec r_{1,2,3}$ in Eq. \ref{dez} as $\beta + \Delta\beta^A$, namely:
\be
A \qquad \vec r_k + \Delta \vec r_k^A \rightarrow \beta + \Delta \beta^A \;.
\ee
We should set this displacement by fitting a datum that corresponds to an intrinsic property of the A-vibration band. Unfortunately, nor the radius of the Hoyle state $\langle r^2\rangle^2_{0^+_2}$, nor the transition rate $B(E2; 2^+_2 \rightarrow 0^+_2)$ are measured (there are however several theoretical calculations).
We show the variation of these two quantities with respect to the extent of vibration in Fig. \ref{bandaA}.
\begin{figure}[!t]
	\includegraphics[width=1.\columnwidth, clip=]{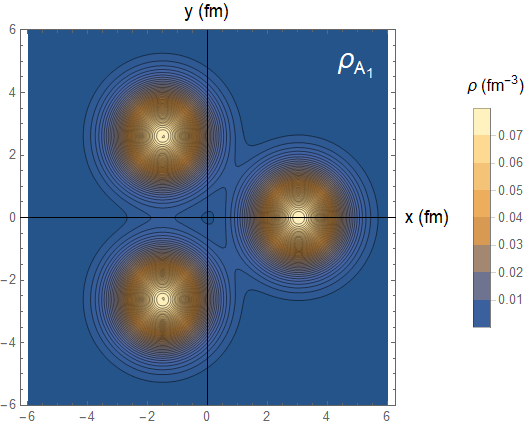}
	\caption{Density of the Hoyle state, that is the first A-type vibration in this model, with the alpha's caught at the moment of maximum elongation.}
	\label{cpA}
\end{figure}
By choosing an intermediate value of $\Delta \beta^A=1.2$ fm that gives a radius of 3.43 fm (about 1 fm more than the g.s. as in Ref. \cite{Ito}) and a transition rate of about 59.6 $e^2$fm$^4$, that is comparable with the calculations of Ref. \cite{Kan}, we can compute the density of the Hoyle state, shown in Fig. \ref{cpA}. The expansion of this density in spherical harmonics, with an expression analogous to Eq. (\ref{dez}), is given in Fig. \ref{rd1}, where one can see the difference of the $\lambda\mu = 00$ term with respect to the ground-state. The central region is depleted and clusterization is more evident.

The transition density connecting the ground state band with the Hoyle band, with $A$ symmetry, can be obtained as an expansion in the small displacements at leading order:
\be
\delta \rho_{gs \rightarrow A} (\vec r) =\chi_1\frac{d}{d\beta}\rho_{gs}(\vec r, \beta)~.
\label{chi1}
\ee
To calculate the transition rates between the g.s. band and the first excited A-type band, one must set the intrinsic transition matrix element $\chi_1$, akin to the parameter used in Ref. \cite{Bij02}, Table  I.  We set $\chi_1=0.247255$ using the value of the monopole matrix element $M(E0)$ in Table \ref{Tab2} fixed at 5.4 $e\; $fm$^2$, that is the value measured in Ref. \cite{Stre,Stre2}. There are other values for this matrix element, namely in \cite{John}, the isoscalar dipole transition is given as an isoscalar energy weighted sum rule strength of $0.08 \pm 0.02 (\%)$.
\begin{figure}[!t]
	\includegraphics[width=1.\columnwidth, clip=]{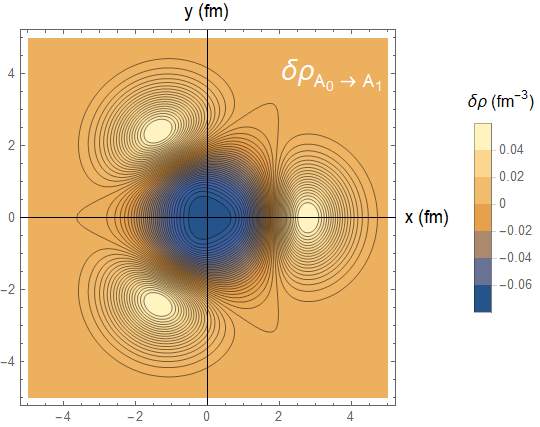}
	\caption{Transition density for the first A-type vibration.}
	\label{cptA}
\end{figure}

\begin{figure}[!t]
	\includegraphics[width=1.\columnwidth, clip=]{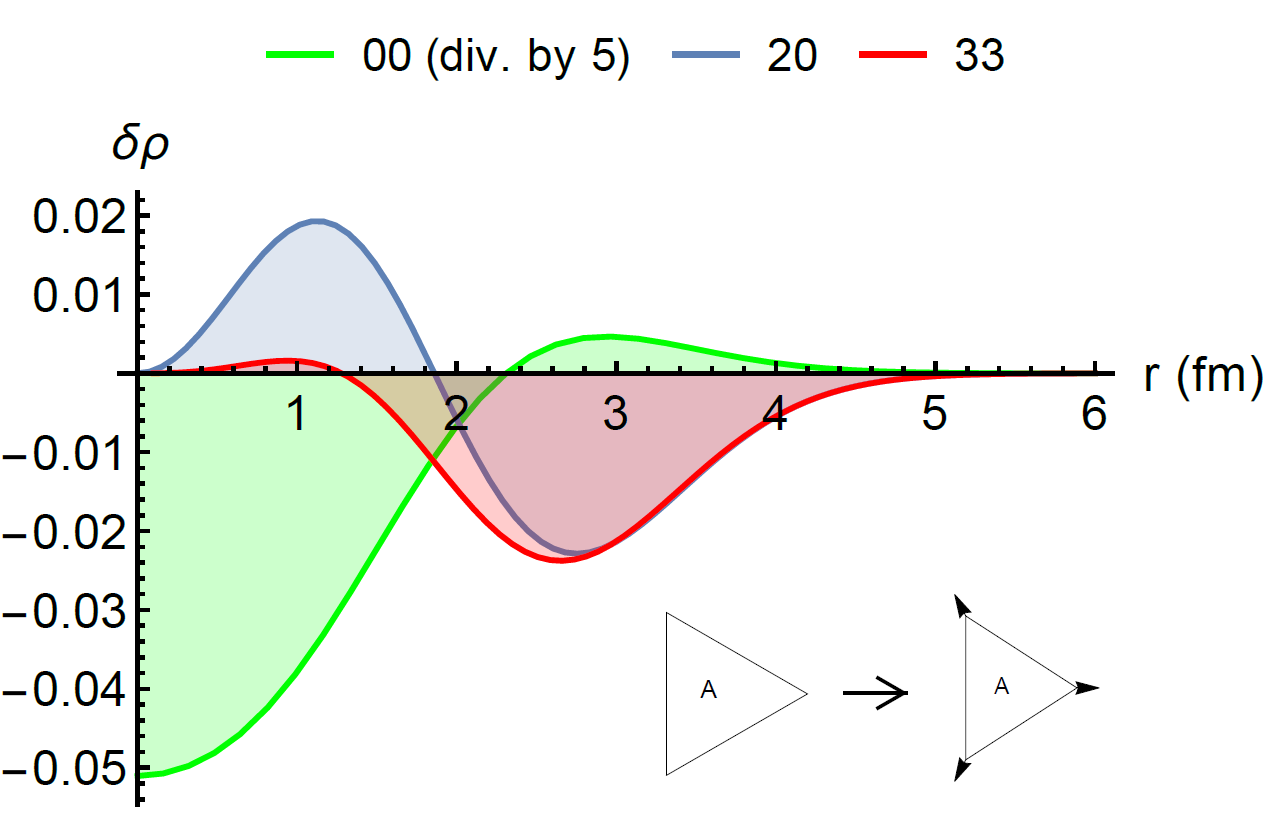}
	\caption{Transition densities for the first A-type vibration and expansion in the lowest order spherical harmonics.}
	\label{tdA}
\end{figure}

\begin{table}[!h]
	\label{Tab2}
	\caption{Quantities calculated in the present work for the Hoyle band, using the values of $\beta$, $\chi_1$  given in the text.}
	\begin{tabular}{lll}\hline \hline
		$\langle r^2 \rangle^{1/2}_{0^+_2}$   &~& 3.44 (fm)\\ \colrule
		$B(E2; 2^+_2 \rightarrow 0^+_1) $ &~&   0.58 (e$^2$fm$^4$)\\ \colrule
		$B(E2; 0^+_2 \rightarrow 2^+_1) $  &~&  2.90 (e$^2$fm$^4$)\\ \colrule
		$B(E3; 3^-_2 \rightarrow 0^+_1) $  &~&  70.42 (e$^2$fm$^6$)\\\colrule
		$M(E0; 0^+_2 \rightarrow 0^+_1) $  &~&   5.4 (e fm$^2$)\\  \botrule
	\end{tabular}
\end{table}
The transition density from the ground-state band to the first excited A-band takes the form:
\be
\delta \rho_{gs \rightarrow A} (\vec r) =\sum_{\lambda\mu}\delta \rho_{gs \rightarrow A}^{\lambda\mu}(r)Y_{\lambda \mu} (\theta,\varphi)
\label{expdrho}
\ee
and it is shown in Fig. \ref{cptA}. The radial components of the expansion in spherical harmonics are shown in Fig. \ref{tdA}, for the allowed values of the projection of the angular momentum $K=0,3,6,\cdots$ \cite{Del17b, Bij02}. The cut in Fig. (\ref{cptA}) shows the moment at which the particles oscillate away from the center in a synchronous fashion, thus depleting the central region (negative transitions density) and enhancing the external regions (positive transition density).
We give in Table \ref{Tab2} the calculated values for other observables.

\begin{figure}[!t]
	\includegraphics[width=1.\columnwidth, clip=]{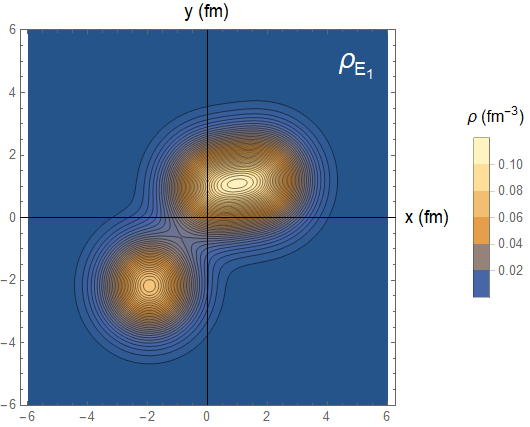}\\
	\includegraphics[width=1.\columnwidth, clip=]{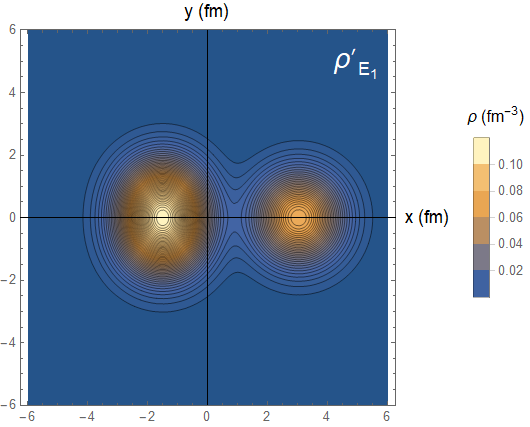}
	\caption{Densities for the first doubly-degenerate E-type vibration. The amplitude of the vibration has been arbitrarily set to $1.2$ fm, the same value used for the A band,  for the sake of illustration.}
	\label{cpE}
\end{figure}

\begin{figure*}[t]
	\includegraphics[width=.32\textwidth, clip=]{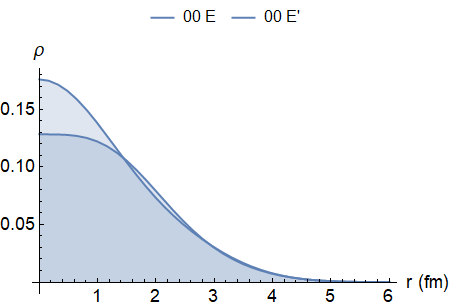}
	\includegraphics[width=.32\textwidth, clip=]{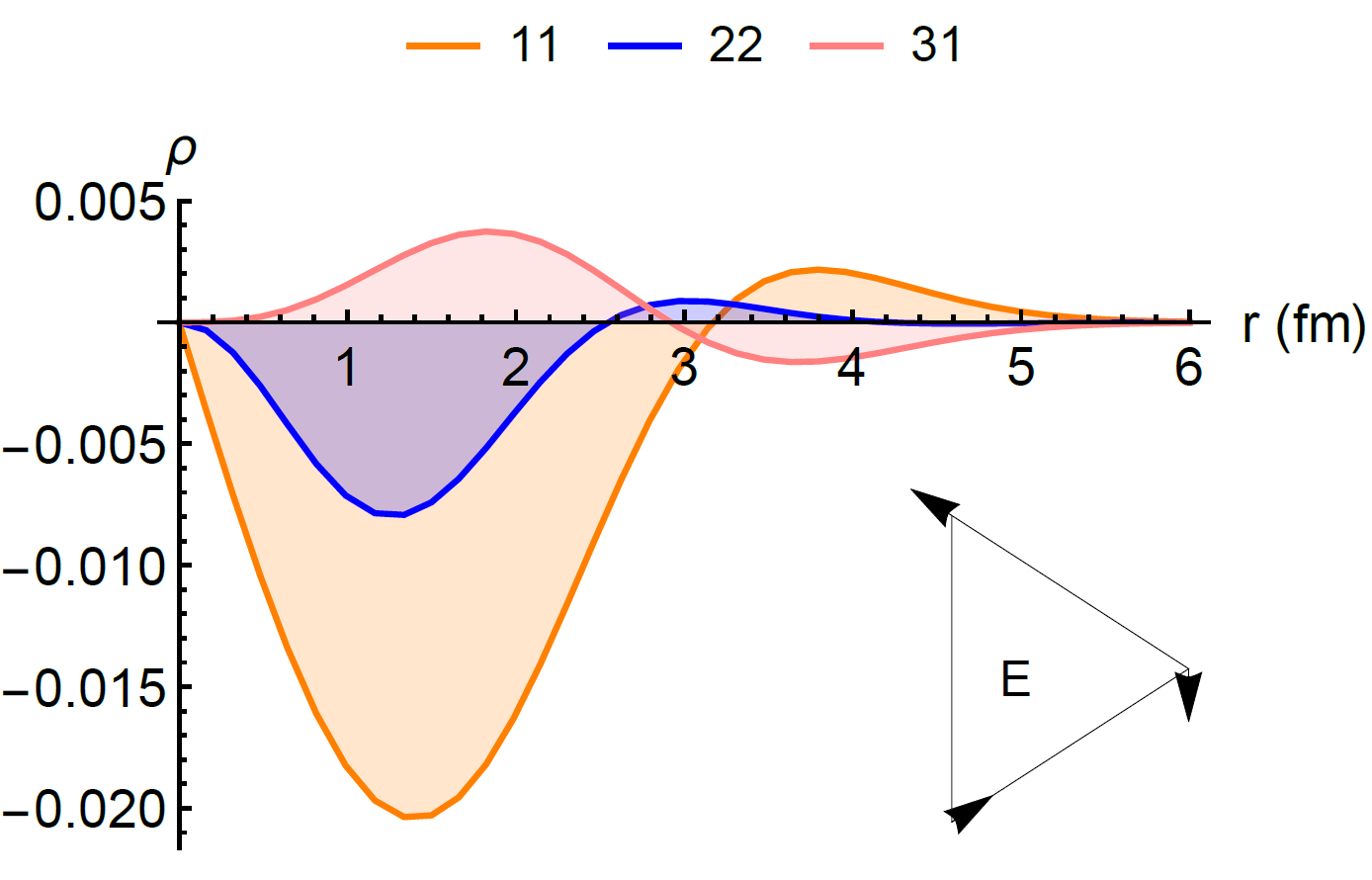}
	\includegraphics[width=.32\textwidth,clip=]{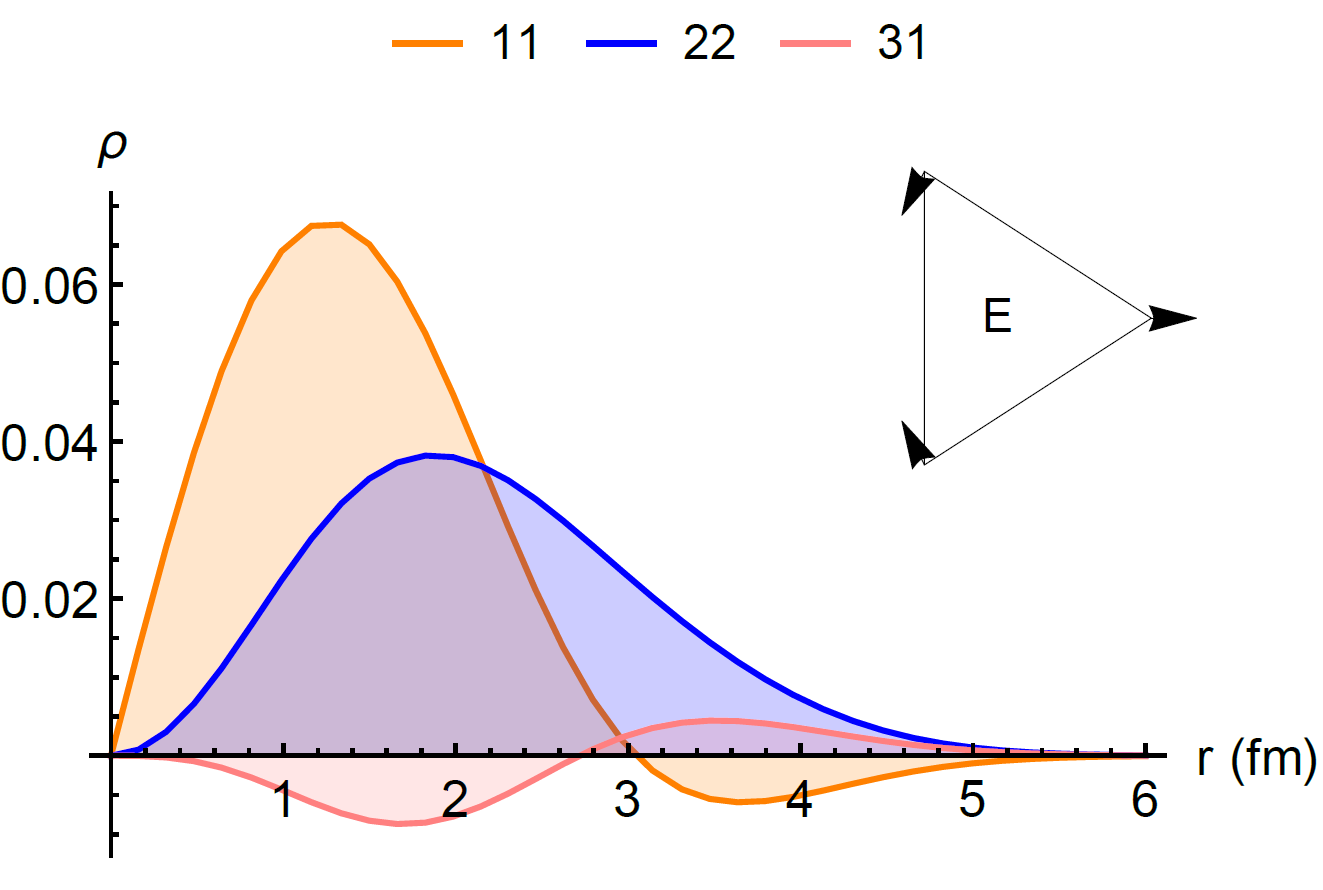} 
	\caption{
		Densities of the two degenerate modes of the E-type vibration. The two dominating $\lambda\mu=00$ components are given in the first panel for both degenerate states, while the others are given separately in the second and third panel. Notice the different vertical scales.}
	\label{cE}
\end{figure*}
\begin{figure}[!h]
\includegraphics[width=1.\columnwidth, clip=]{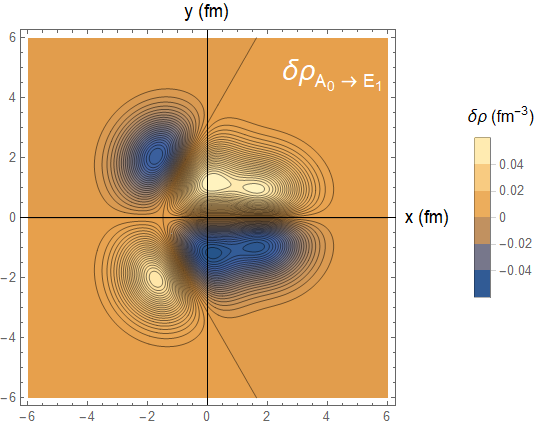} \\
\includegraphics[width=1.\columnwidth, clip=]{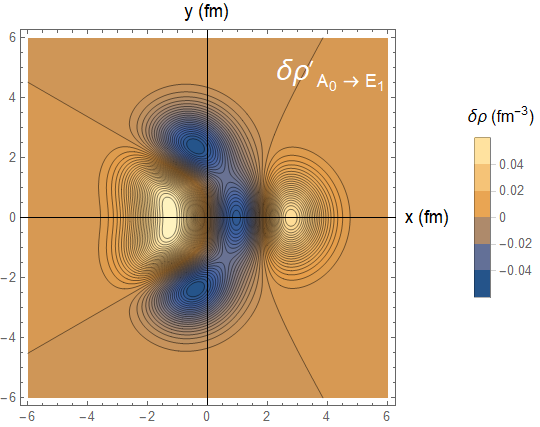} 
\caption{
 The doubly degenerate E-type vibrations have two transition densities $\delta\rho$ corresponding to the two normal modes of motions shown in the inset of Fig. (\ref{tdE}).}
\label{tdE}
\end{figure}

Together with the A-type normal mode, there is another doubly-degenerate normal mode with $E$ symmetry. The two panels of Fig.\ref{cpE} show the densities of the doubly-degenerate E-type band, whereas the panels of Fig. \ref{tdE} show the corresponding transition densities. 
The vector displacements associated with this mode are shown in the insets of Fig. \ref{cE}.
This picture shows the expansion of the densities of the two degenerate components of the $1_1^-$ state.
Now one cannot simplify the notation to the radial variable only as in the preceding case, because these vectors do not point along the radial direction, therefore in principle one should take:
\be
E \qquad \vec r_k + \Delta \vec r_k^E  \qquad  \mid  \Delta \vec r_k^E\mid \rightarrow \eta\;,
\ee
but since the direction of the vectors is fixed, we will consider only the magnitude of the displacements, that we call $\eta$. 
In Fig. \ref{cpE}, the amplitude of the vibration has been arbitrarily set to $1.2$ fm, i.e. the same value that was used in the A-band to set a root mean square radius about 1 fm larger than in the g.s., for the sake of illustrating the fact that this vibration correspond to the channel forming $^8$Be plus an $\alpha$ particle: in fact, in both plots, one of the alpha's retains its density almost intact and detaches from the rest. With this choice, the root mean square radius for the E-band is intermediate between the g.s. and Hoyle values.

The intrinsic transition density from the grous to the E band takes a form similar to Eq. (\ref{chi1}), namely
\be
\delta \rho_{gs \rightarrow E} (\vec r) =\chi_2\frac{d}{d \eta}\rho_{gs}(\vec r, \eta)~.
\ee
where the value of $\chi_2$ should be set using some experimental observable. This is difficult to be accomplished here, as the only information easily accessible is the model-dependent isoscalar dipole matrix element value given in Ref. \cite{Kan, John} 
and extracted in $\alpha-$scattering experiments. In the present case we adopt $\chi_2= 0.136$ and obtain the value $M(IS1; 0_1^+ \rightarrow 1_1^-) \simeq 0.31 e $fm$^3$.
The isoscalar dipole transitions are calculated using the definition of the matrix element as:
\begin{equation}
M(IS 1; 0^+_1 \rightarrow 1^-)= Z \int  \delta\rho_{gs\rightarrow E}^{1} (r) \Bigl(r^3-\frac{5}{3}\langle r^2\rangle r\Bigr) r^2 dr
\end{equation}

\begin{figure}[!h]
\includegraphics[width=1.\columnwidth, clip=]{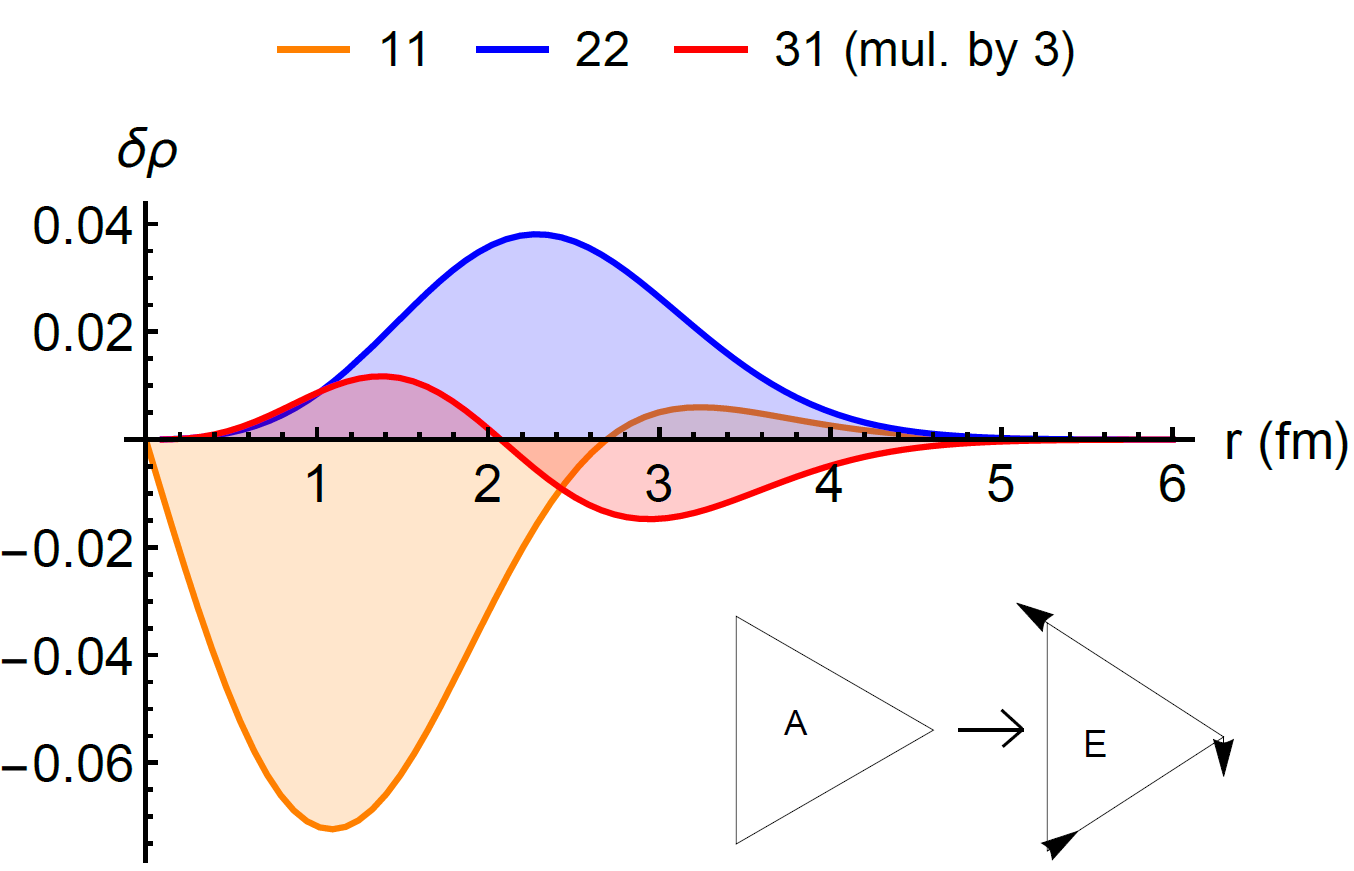} 
\caption{Radial transition densities $\delta\rho_{gs \rightarrow E}^{\lambda \mu} (r)$ to the E-type band for the first type of motion (cfr. inset). The corresponding transition densities for the second type of motion can be obtained upon changing sign for $11$ and $31$.}
\label{tdEr}
\end{figure}

The transition densities for the E-type vibrations are expanded in multipoles
\be
\delta \rho_{gs\rightarrow E} (\vec r) =\sum_{\lambda\mu}\delta \rho_{gs\rightarrow E}^{\lambda\mu}(r)Y_{\lambda\mu} (\theta,\varphi) 
\ee
and the radial part of the transition densities for the first few states, having $K=1$ or $K=2$ (and, more in general, all values of $K$ that are not divisible by 3) are shown in Fig. \ref{tdEr}. Notice that the curves are the same for the two degenerate modes. The smaller one, i.e. the $\{\lambda\mu\}=\{31\}$ component, has been magnified three times to make it comparable with the others.

\section{Form factors}
\begin{figure}[h]
	\includegraphics[width=1.\columnwidth,clip=]{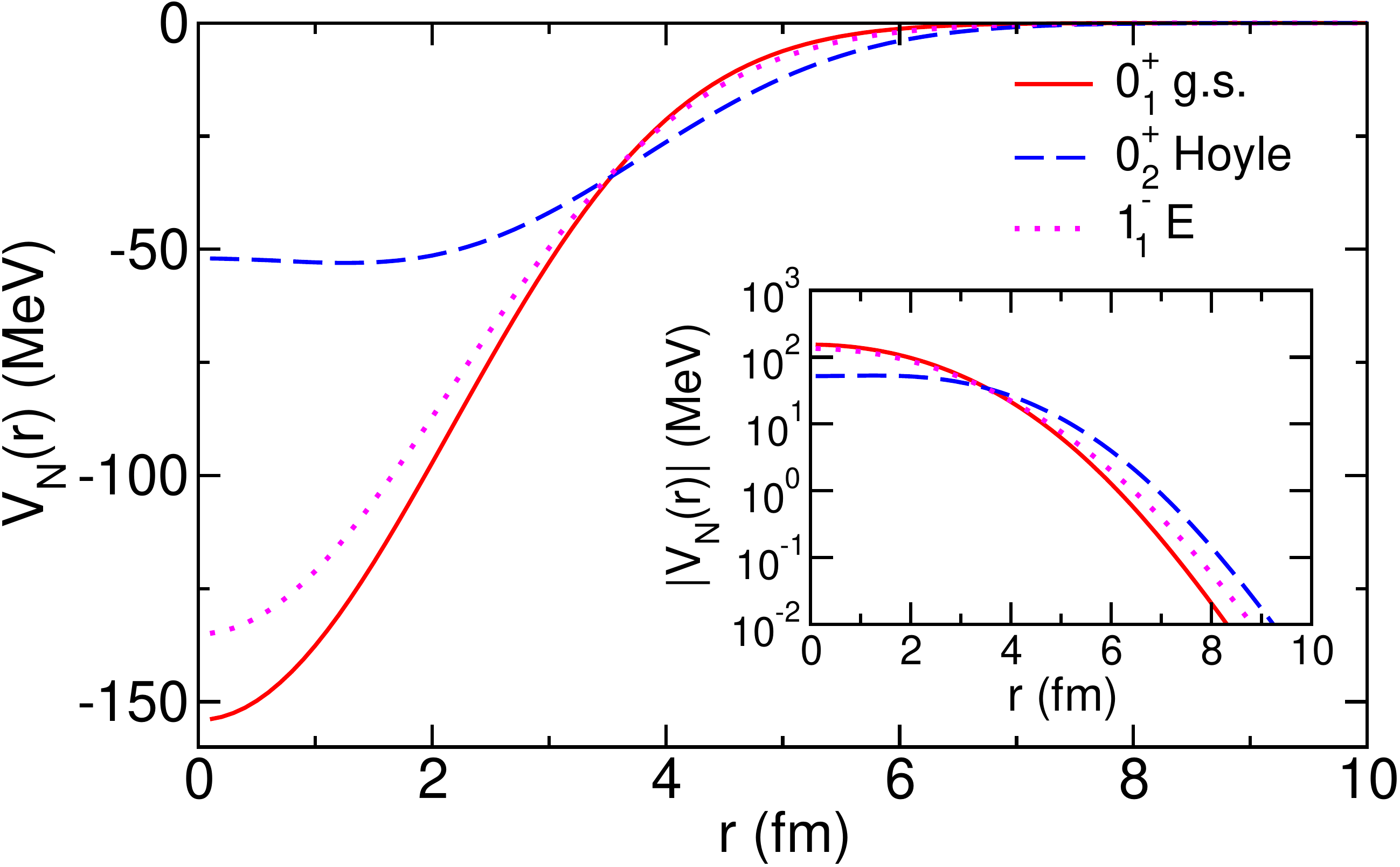}
	\caption{Double folding nuclear potentials for the system $\alpha$ +$^{12}$C for the ground (red dot-dashed line), the "Hoyle" (blue dashed) and the E-type (dotted) band states. The inset shows the same quantity (absolute values) in logarithmic scale to appreciate the different radial extension on the tails.}
		\label{dfpot}
	\end{figure}

The densities and transitions densities described above in the equilateral triangular cluster model contain all the structure information to compute form factors for inelastic excitation processes such as the $\alpha + ~^{12}$C scattering, provided one chooses a suitable nucleon-nucleon potential. We construct the real part of the nuclear optical model potential using a double-folding prescription as in Ref. \cite{sat, sat1}, namely
\be
V_{N}(R)~=~\int\int \rho_{\alpha}(\vec r_1-\vec R) ~\rho_{T}(\vec r_2)~ v_{N}(r_{12})~ d{\vec r_1} d{\vec r_2}
\ee
where $\rho_{\alpha,T}$ are the densities of projectile and target and the effective interaction $v_N$ is a function of the nucleon-nucleon distance $r_{12}$. In this case the $\alpha$ particle is an isoscalar probe ($N=Z$ system), therefore only the isoscalar part of the interaction will contribute to the integral. The widely used density dependent Reid type M3Y nucleon-nucleon interaction is used for $v_N$ \cite{m3y, sat}. Of course, due to the density dependence, the folding potential is different for each combination of projectile and target densities: in our case the $\alpha$ particle is always in the ground state for low-energies, while the target can be in any of the selected low-energy states. The potential might differ significantly in the interior, but is quite similar on the surface region, as shown in Fig. (\ref{dfpot}), that is the relevant one for grazing processes. The potential in the case of the Hoyle state has a slightly longer tail, owing to the fact that the densities of Fig. \ref{cpA} show indeed a larger range with respect to the ground state. In the figure we give also the potential used in the case of the two degenerate $1^-$ states, that are almost identical. The inset shows the same potentials in logarithmic scale to appreciate the differences on the tail.

\begin{figure}[h]
		\includegraphics[width=0.5\textwidth,clip=]{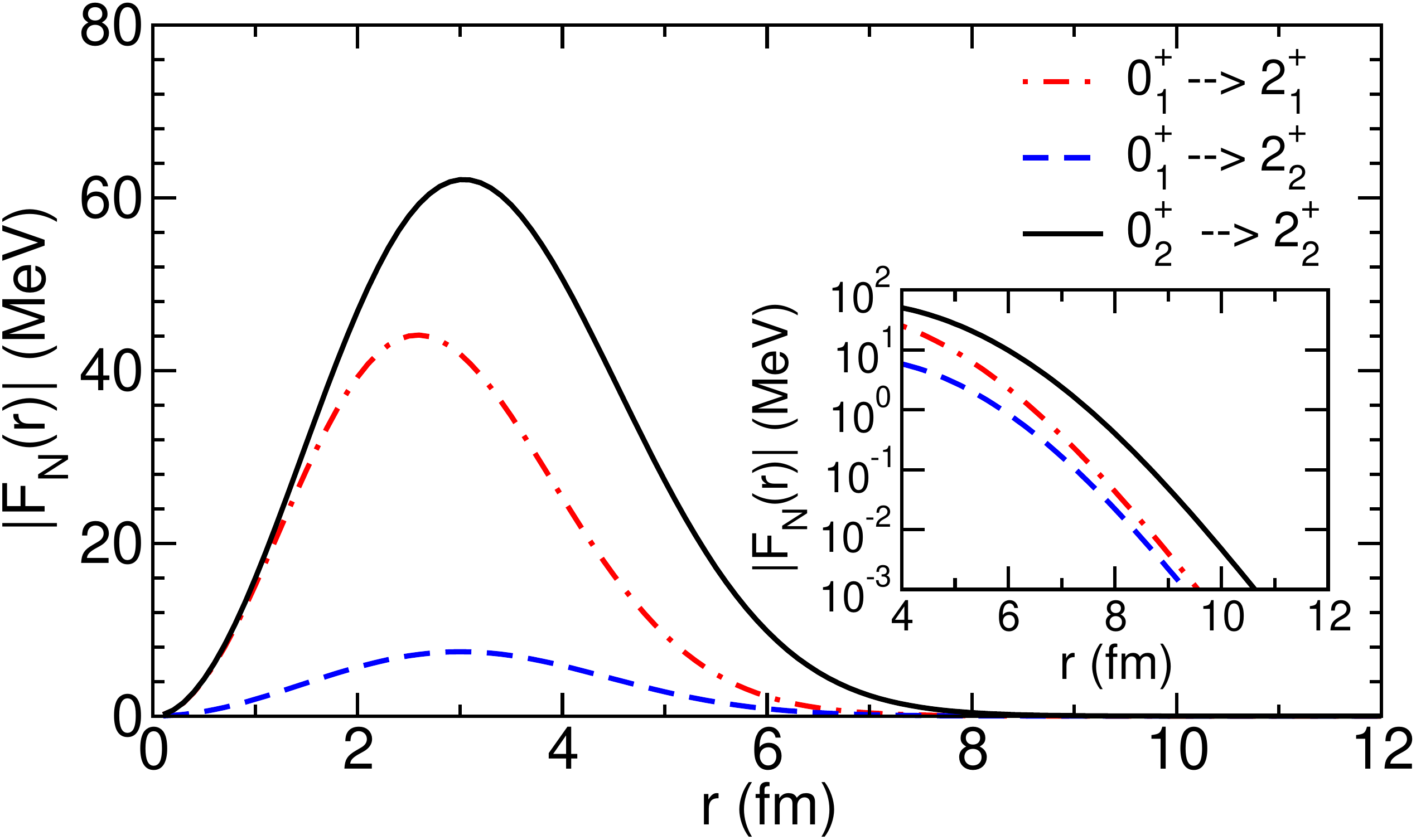}
		\caption{Absolute value of radial form factor for the system $\alpha$ +$^{12}$C for the three excitation processes shown in the legend. The quadrupole states are the one built on top of the ground state (2$^+_1$) and on top of the 0$^+_2$ "Hoyle" state (2$^+_2$)}
		\label{FFlin}
	\end{figure}

\begin{figure}[h]
		\includegraphics[width=1.\columnwidth,clip=]{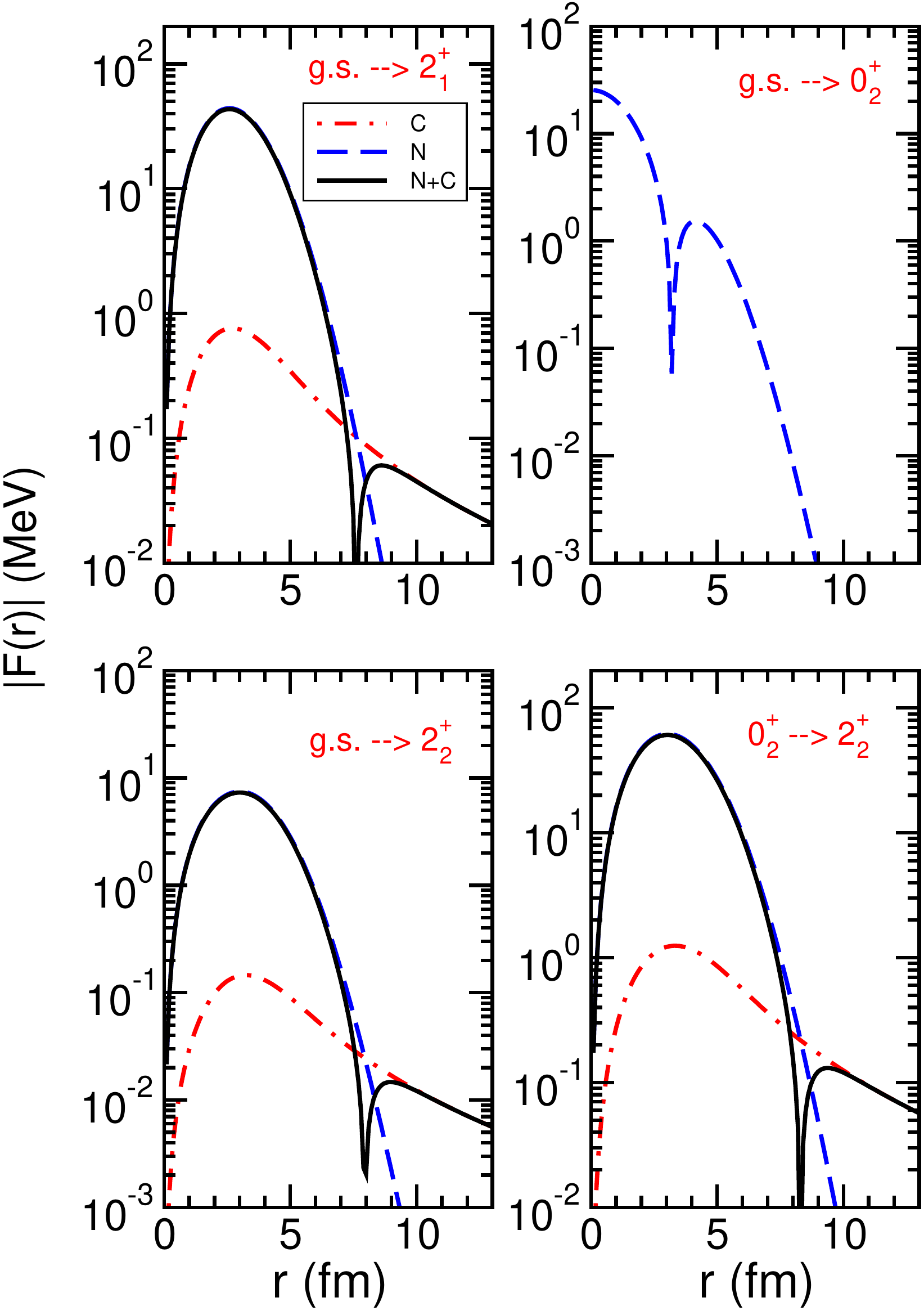}
		\caption{Form factors in logarithmic scale for a few inelastic excitation processes of interest. We show the nuclear, coulomb and total form factors.}
		\label{FFlog}
	\end{figure}

Using the transition densities calculated above, one can also compute non-diagonal matrix elements and calculate the form factors by double-folding:
$$F_{ij}(\vec R)= F_{ij}(R) Y_{\lambda \mu}(\hat R) = $$
\be
=\int\int \rho_{\alpha}(\vec r_1-\vec R)~v(r_{12})~ \delta\rho_{i\rightarrow j}(\vec r_2)~ d{\vec r_1} d{\vec r_2}~
\ee
where $v$ contains the nuclear and coulomb interactions.
We show in Fig. \ref{FFlin} a few lowest form factors in linear scale to appreciate the difference in magnitude and range, and we show in Fig. \ref{FFlog} a compilation of form factors in logarithmic scale, where the nuclear and Coulomb contributions are shown together with the total. Clearly the 0$^+_{gs} \rightarrow $ 0$^+_2$ monopole transition has only the nuclear part. 
	
\begin{figure}[t!]
	\includegraphics[width=1.\columnwidth,clip=]{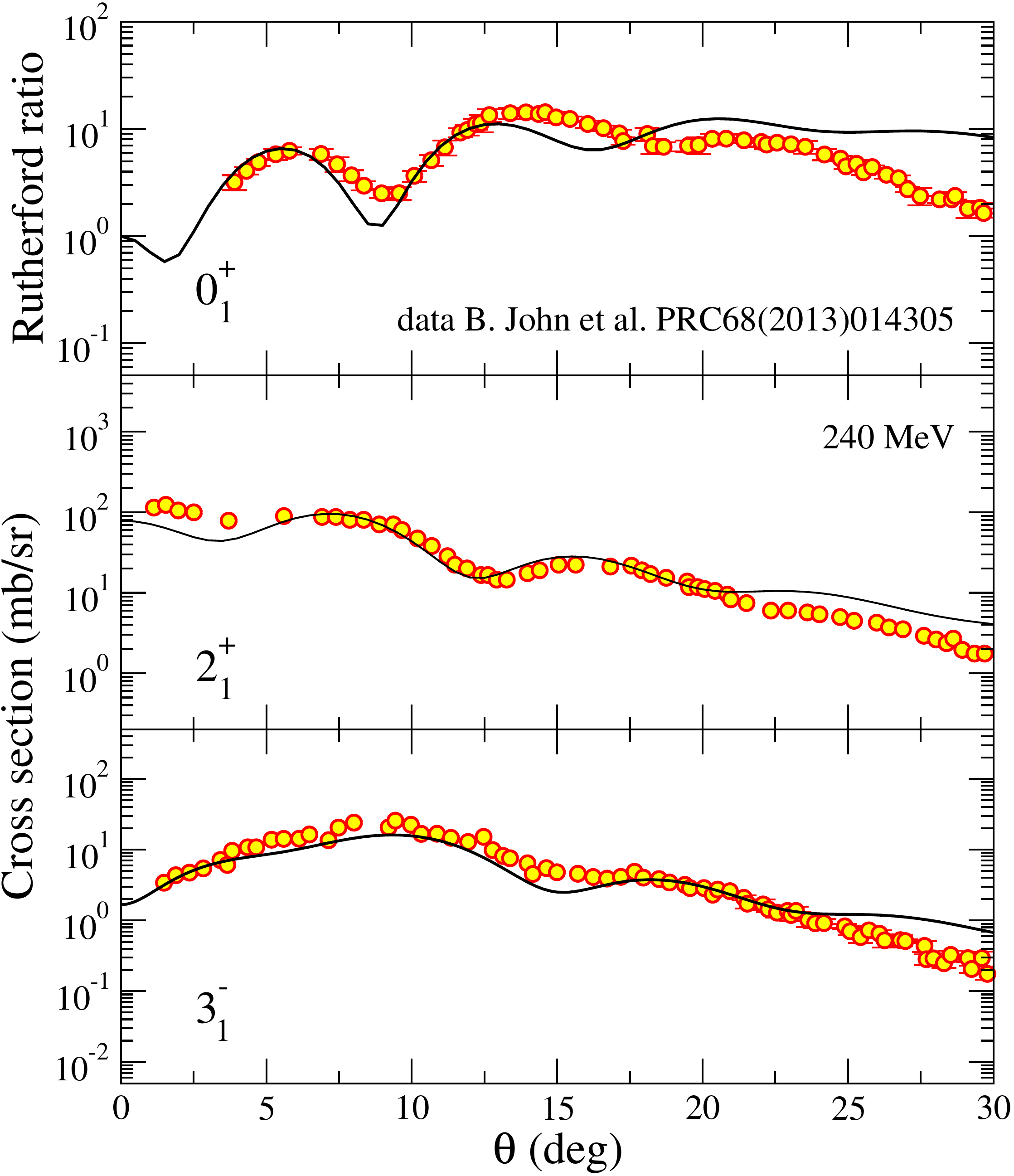}
	\caption{Differential cross-section for the elastic scattering and the transitions $0_1^+ \rightarrow 2_1^+$ and $0_1^+ \rightarrow 3_1^+$ at 240 MeV bombarding energy. Data are from Ref. \protect\cite{John} (retrieved through EXFOR). }
	\label{dwba_01-21}
\end{figure}
In these figures the excitation processes of interest are those related to the 2$^+$ states in the ground and ``Hoyle" bands. The comparison between the intra- and inter-band transitions shows  that the form factor for the transition from 0$^+_2$ to 2$^+_2$ has a larger radial extension than the other two transitions taken into consideration. As a consequence the angular distribution for this transition may cover a reduced angular range compared to the other ones and therefore might give a hint on the radial extension of the 2$^+_2$ state as pointed out in Ref. \cite{Ito}.
The strong inband coupling could give rise instead to a significant interference between the direct population of the  2$^+_2$ state and the two-step process via the 0$^+_2$ state. This interference, once plugged into a coupled-channel calculation could give information on the different radial size of the ground and Hoyle bands as a function of scattering angle and bombarding energy. 

\begin{figure}[t!]
	\includegraphics[width=1.\columnwidth,clip=]{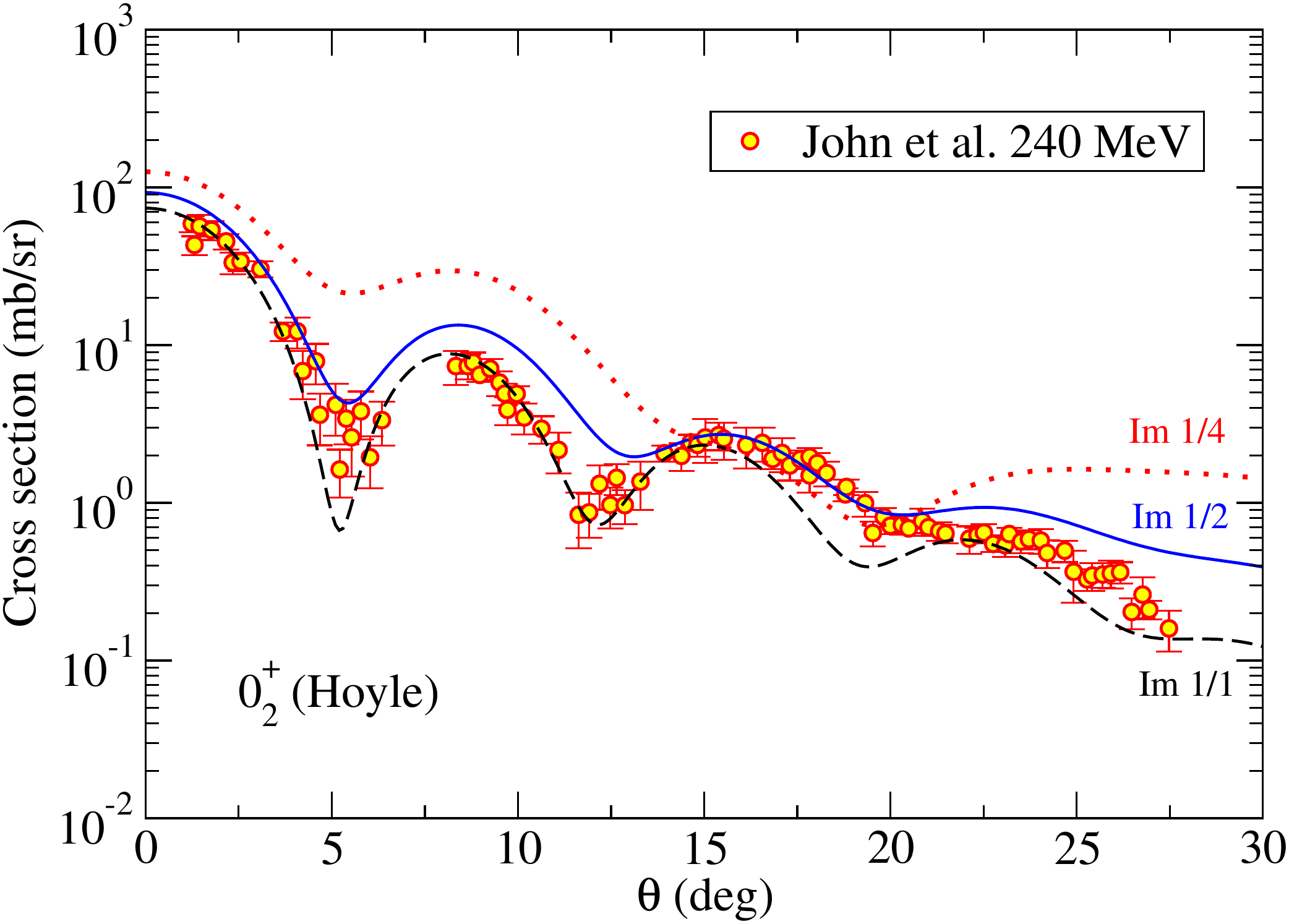}
	\caption{Differential cross-section for the transition $0_1^+ \rightarrow 0_2^+$ at 240 MeV bombarding energy. Data are from Ref. \protect\cite{John} (retrieved through EXFOR) and the three curves have different factors for the depth of the imaginary part as indicated in the figure.}
	\label{dwba_01-02}
\end{figure}
In order to illustrate that the simple geometrical model is able to capture the main features in reactions where alpha-cluster degrees of freedom are involved, we have computed DWBA differential cross-sections for $\alpha+~^{12}$C inelastic scattering.
In Fig. \ref{dwba_01-21} we show the differential cross-section (or ratio to Rutherford in the elastic case) for the ground-state elastic scattering and for the excitation to the first excited $2^+$ and $3^-$ states for the $\alpha+~^{12}$C reaction. In these calculations we have set the imaginary part of the optical potentials and form factors to $1/2$ of the real part. The experimental data \cite{John} are retrieved through the EXFOR database. Among the many possible bombarding energies, we have taken the dataset at 240 MeV as an example of the ability of the alpha-cluster model to reproduce correctly the shape and magnitude of these cross-sections. There is some deviation of the calculated line with respect to the data at angles above the grazing angle, but the overall behaviour is well reproduced. These results are encouraging and indicate that, despite the complications that one might invoke in other models, the present simple approach is enough to explain the data, a symptom that alpha clustering plays a vital role here. In Fig. \ref{dwba_01-02}, we give the calculations and data for the  $0_1^+ \rightarrow 0_2^+$ transition (ground to Hoyle): in this case we have explored the sensitivity of the DWBA calculations to the depth of the imaginary part of the potential. The three curves (dotted red, solid blue and dashed black) correspond to the imaginary part set to 1/4, 1/2 and equal to the depth of the real part. The shape of the curve is again very good and clearly the best agreement is found for values between 1/2 and 1. This is an indication that not only the description of the ground state band is good, but also the description of the Hoyle band in terms of breathing vibration finds confirmation in reaction data.

\begin{figure}[t!]
	\includegraphics[width=1.\columnwidth,clip=]{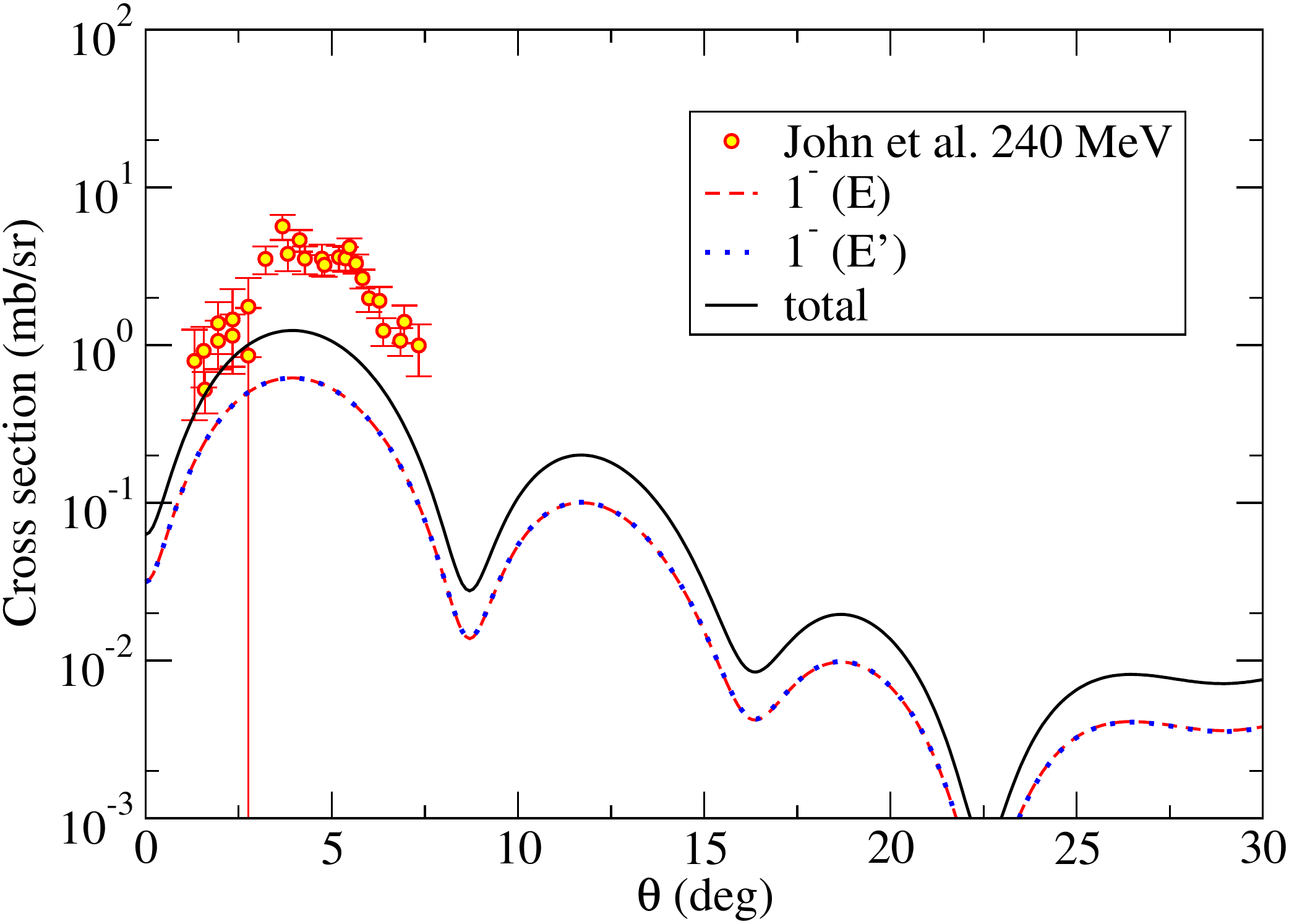}
	\caption{Differential cross-section for the transition $0_1^+ \rightarrow 1_1^-$ at 240 MeV bombarding energy. Data are from Ref. \protect\cite{John} (retrieved through EXFOR).  The dashed colored curves show the cross-section to the two degenerate components, indicated with E and E' and the black solid curve shows the sum.}
	\label{dwba_01-11}
\end{figure}

In Fig. \ref{dwba_01-11} we show the differential cross-section to the bandhead of the E-type band, that is the sum of two components. Despite the geometrical differences, the averaged integrals give the same results for the two components. The comparison with data is good at very small angles, but fails to reproduce the peak at around 4-5 degrees. Clearly, this  depends on the choice of the $\chi_2$ parameter, that we had previously fixed to the matrix element of the model-dependent isoscalar electric dipole transition. A change by a factor of two (that would take $\chi_2$ to about the same value of $\chi_1$) would reproduce the peak and slightly overestimate the cross-section at the smallest angles. We have checked that a change of $\eta$ within the range $[1.0, \cdots, 1.4]$ does not affect appreciably the final result. Instead, a change in the imaginary part of the potential does change the differential cross-section to the $1_1^-$ state, but mostly at larger scattering angles. The change within the extent of the data is not very relevant.

We have confined ourselves to DWBA calculations in the present paper, for the sake of illustrating the validity of the approach. Coupled channel effects may show up if one includes higher excited states, as shown in Ref. \cite{KanOg}, but are not necessary to the present discussion. For instance, our DWBA calculations do not reproduce the data at angles around 5 degrees in Fig. (\ref{dwba_01-11}), while they give a reasonable reproduction at smaller angles. The same is observed in Ref. \cite{Kan}, where, in addition, coupled channel calculations are performed that better fit the existing data. Clearly a simple DWBA approach, based only on the direct transitions from the ground state cannot be satisfactory in presence of strong second order couplings.
A typical example is the population of the $2^+_2$ state in the Hoyle band, where due to the strong $B(E2; 0^^+_2 \rightarrow 2^+_2)$, the direct transition competes with the strong two-step process passing through the $0^+_2$ state \cite{Ito,KanOg}.

\section{Conclusions}
We have investigated the performance of the alpha-cluster model for $^{12}$C in reproducing structure and reaction observables. We have set up the densities of the alpha cluster and the carbon-12 nucleus as an equilateral triangular arrangement according to the recent prescriptions used in the Algebraic Cluster Model. We have constructed transition densities to the states of the ground band and to the states of excited $A$ and $E$ bands. In doing so, we have chosen some parameters to reproduce a minimal set of known matrix elements or electromagnetic transition rates, and we have seen that the model is able to give a quantitative interpretation of almost all the available data. 
The densities have been used to generate double folding potentials between the alpha particle and the $^{12}$C nucleus. Form factors for several transitions have been computed and used in DWBA calculations to show that this molecular model with a very simple geometry is indeed able to reproduce the shape and magnitude of many scattering data. Notably, the ground to Hoyle  $0_1^+ \rightarrow 0_2^+$ transition, that involves a static equilateral triangular configuration in the ground state and an oscillating equilateral triangle breathing mode for the Hoyle state, provides a satisfactory explanation for the scattering data, if one allows for a slightly deeper than usual imaginary part of the optical potential. We have also investigated the cross-section to the doubly-degenerate E-band. Despite the quite significant geometrical differences between the two components of Fig. \ref{cpE}, their form factors and reaction observables look practically the same. 

In conclusion, alpha clusters, if properly described into a fully quantal molecular approach, not only play a role in the Hoyle state, as it was commonly believed until a few years ago, but they are strongly involved into the structure of the ground state and in a large part of the lowest excited states. The role of the fermionic degrees of freedom and the Pauli principle does not seem to be crucial for the description of structure and reactions where the cluster degrees of freedom are involved. In addition to the structure properties, this model can be effectively applied to reaction observables, thus significantly enlarging the amount of data that can be described in the molecular approach that includes rotations and vibrations of a simple triangular cluster configuration.

It is worth noting that the present approach does not take into account the unbound nature of $^{12}$C states above three-$\alpha$ threshold, starting with the Hoyle state. Three-body calculations with $\alpha$-$\alpha$ scalar interactions, including continuum effects explicitly, could provide more insight into the structure properties of this nucleus. Work along these lines is ongoing and will be presented elsewhere, together with full coupled-channels calculations for the $\alpha+{^{12}}$C inelastic excitation within both approaches, which is definitely essential in the population of e.g. the $2^+_2$ state.

\section*{Appendix}
The states in the laboratory frame can be written as

$$
\mid IM, n_A n_E \rangle =$$
\be\sum_K \underbrace{ \sqrt{\frac{2I+1}{16\pi^2(1+\delta_{K,0})}}}_{{\cal N}_K} \bigl(\WD^{(I)*}_{MK} +(-1)^K \WD^{(I)*}_{M,-K} \bigr)\mid n_A n_E\rangle
\ee
where the intrinsic state $| n_A n_E \rangle$ is labeled by the number of phonons of each type, such that the ground state is $| 00 \rangle$,while the bandheads of the $A$ and $E$ type first vibrations are $| 10 \rangle$ and $| 01\rangle$ respectively.

The transition density in the laboratory frame can be related to that into the intrinsic frame with
$$
\langle I_f M_f, n_A n_E \mid \hat \rho \mid  I_i M_i, n_A n_E  \rangle =$$
\be
 \sum_{\lambda\mu} \delta \rho_{\lambda\mu} (r) \sum_{K_i,K_f} {\cal N}_{K_f}^* {\cal N}_{K_i} \sum_\kappa C Y_{\lambda\kappa} (\theta\varphi)
\ee 
where the summations are taken on non-negative values of $K$'s and where
$$
C=\frac{8\pi^2}{(2I_f+1)} \langle I_f\lambda I_i\mid M_f \mu M_i\rangle \Bigl(  \langle I_f\lambda I_i\mid K_f \kappa K_i\rangle + $$
$$(-1)^{K_f}\langle I_f\lambda I_i\mid (-K_f) \kappa K_i\rangle+(-1)^{K_i}\langle I_f\lambda I_i\mid K_f \kappa (-K_i)\rangle+$$
\be
+(-1)^{K_f+K_i}\langle I_f\lambda I_i\mid (-K_f) \kappa (-K_i)\rangle\Bigr) 
\ee

From this, one can calculate reduced matrix elements and probabilities.

\end{document}